\documentclass[doublecol]{epl2}

\usepackage{amsmath,amssymb}
\renewcommand{\le}{\leqslant}
\renewcommand{\ge}{\geqslant}
\renewcommand{\vec}[1]{{\boldsymbol{#1}}}

\newcommand{\defeq}{:=}
\newcommand{\cc}{c_{\mathrm{crit}}}
\newcommand{\Rg}{R_{\mathrm{gyr}}}
\newcommand{\df}{d_{\mathrm{f}}}
\newcommand{\dfs}{d_{\kern0.5pt\mathrm{f},\kern0.5pt\mathrm{s}}}
\newcommand{\UEV}{U_{\mathrm{EV}}}
\newcommand{\eq}{\mathrm{eq}}
\newcommand{\stat}{\mathrm{stat}}

\title{Variational bounds for the shear viscosity of gelling melts}

\author{Claas H.\ K\"ohler,\inst{1} Henning L\"owe,\inst{2} 
  Peter M\"uller\inst{1} \and Annette Zippelius\inst{1}}

\shortauthor{C. H. K\"ohler, H. L\"owe, P. M\"uller, A. Zippelius}

\institute{%
  \inst{1} Institut f\"ur Theoretische Physik,
  Georg-August-Universit\"at, 37077 G\"ottingen, Germany\\ 
  \inst{2} Swiss Federal Institute for Snow and Avalanche Research,
    Fl\"uelastr.\ 11, CH--7260 Davos Dorf, Switzerland
} 

\pacs{64.60.Ht}{Dynamic critical phenomena}
\pacs{61.25.Hq}{Macromolecular and polymer solutions; polymer melts; swelling}
\pacs{61.20.Lc}{Time-dependent properties; relaxation}

\abstract{We study shear stress relaxation for a gelling melt of
  randomly crosslinked, interacting monomers. We derive a lower bound
  for the static shear viscosity $\eta$, which implies that it
  diverges algebraically with a critical exponent $k\ge 2\nu-\beta$.
  Here, $\nu$ and $\beta$ are the critical exponents of percolation
  theory for the correlation length and the gel fraction.  In
  particular, the divergence is stronger than in the Rouse model,
  proving the relevance of excluded-volume interactions for the
  dynamic critical behaviour at the gel transition. Precisely at the
  critical point, our exact results imply a Mark--Houwink relation for
  the shear viscosity of isolated clusters of fixed size.}

\begin{document}

\maketitle

\section{Introduction}

The most prominent feature of a gelling (macro-) molecular liquid is
the divergence of the static shear viscosity as the gelation
transition is approached from the sol side. The melt becomes
increasingly more viscous, reflecting larger and larger solid like
regions and signalling the emergence of shear rigidity in the
amorphous solid state of the gel. Many experiments have tried to
measure the divergence of the shear viscosity quantitatively, yet the
data are still very confusing.

The gelation transition is known to be a percolation transition, which
is controlled by the crosslink density $c$, see e.g.\ \cite{StCo82}
for a review. It has been analysed with the tools of critical
phenomena, in particular it has been verified that the cluster size
distribution follows the scaling law $ f(n) \sim n^{-\tau}
\exp\{-n/n^{*}\}$ if the distance to the critical point, $\cc$, is
sufficiently small. Here $n$ is the cluster size or more precisely the
number of crosslinked monomers of a cluster. The quantity $n^{*} \sim
|c-\cc|^{-1/\sigma}$ has the meaning of a ``cut-off size'' and
$\sigma$ and $\tau$ are critical exponents \cite{StAh94}.

The divergence of the static shear viscosity is usually analysed in
terms of an algebraic singularity $\eta\sim (\cc -c)^{-k}$ and a
dynamic critical exponent $k$. However the values of $k$ which have
been deduced from experiment vary enormously -- in
\emph{contradiction} to the concept of \emph{universal exponents} put
forward by the dynamic renormalisation group theory. To appreciate the
scatter of the data we summarise the reported values for $k$ in
Table~\ref{data}. We have also included results from simulations in
Table~\ref{data}, based either on the bond-fluctuation
model\cite{GaAr00} or on molecular dynamics in combination with a soft
pair potential\cite{VePl01,JePl03}.

\begin{largetable}
  \caption{Numerical values for the viscosity exponent from
    experiments and simulations}
  \medskip
  \label{data}
  \begin{tabular}{c||c|c|c|c|c|c|c|c|c|c|c|c|c|c|c|c}
    Ref. & \cite{AdDe79} & \cite{AdDe85} & \cite{MaAd88} & \cite{AxKo90} &
    \cite{TaUr90} & \cite{CoGi93} & \cite{DeBo93} & \cite{TaYo94} &
    \cite{LuMo95} & \cite{ZhTh96} & \cite{AdLa97} & \cite{LuMo99} &
    \cite{ToFa01}& \cite{VePl01}& \cite{JePl03}& \cite{GaAr00} \\
    \hline
    $k$ & 0.78 & 0.81 & 1.4 & 0.82 & 0.2 & $>$1.4 & 0.7 & 1.3 & 1.36 & 1.27 &
    1.1 & 6.1 & 0.76 & 0.7 & 0.65 & 1.3
  \end{tabular}
\end{largetable}

The origin of the broad scatter of the data is not understood at
present. It has been suggested that there might be more than one
universality class for the dynamic critical exponents. For example, in
dense solutions hydrodynamic interactions play a crucial role, whereas
in melts such interactions are absent. However, it is not obvious, how
to group the above data into two universality classes. Another
explanation for the scatter has evoked the size of the critical
region, which shrinks with growing linear size of the crosslinked
(macro-) molecules\cite{DeGe77}.  Consequently some experiments could
be outside the critical region or measure crossover exponents.

In view of the controversial results it is highly desirable to obtain
rigorous results for the divergence of the static shear viscosity. In
this paper we consider a Rouse model, generalised to account for
excluded-volume interactions, and derive an exact lower bound for
$\eta$. This allows us to conclude that the viscosity diverges at
least as strongly as the lower bound, implying the inequality $k\ge
2\nu-\beta$ for the critical exponent $k$. Here, $\nu$ and $\beta$ are
the critical exponents of percolation theory for the correlation
length and the gel fraction. In $d=3$ spatial dimensions the lower
bound for the exponent implies $k \ge 1.32$, in disagreement with
several entries in Table~\ref{data}. Furthermore the lower bound
exceeds the value for the critical exponent of the Rouse model without
excluded volume, proving that excluded-volume interactions are indeed
relevant for the dynamic critical behaviour.  We also derive a lower
bound for the case of a \emph{single} (isolated) cluster with $n$
monomers.  This lower bound is found to agree with an upper bound,
resulting in an exact expression for the shear viscosity
$\eta_{n,\,\mathrm{s}}$ for single clusters of fixed size $n$.  If the
radius of gyration scales with cluster size according to $\Rg\propto
n^{1/\dfs}$, where $\dfs$ is the Hausdorff
fractal dimension for isolated clusters, then the single-cluster
viscosity is shown to obey the Mark--Houwink scaling relation
$\eta_{n,\,\mathrm{s}}\sim n^{2/\dfs}$.

%%%%%%%%%%%%%%%%%%%%%%%%%%%%%%%%%%%%%%%%%%%%%%%%%%%%%%%%%%%%%%%%%%%%%%%
%
\section{Model}
\label{model}
%
%%%%%%%%%%%%%%%%%%%%%%%%%%%%%%%%%%%%%%%%%%%%%%%%%%%%%%%%%%%%%%%%%%%%%%%

We consider a system of $N$ identical particles at positions $\{\vec{R}_1,
\ldots,\vec{R}_N\}$ in a $d$-dimensional volume $V$. Permanent
crosslinks connect $M$ randomly chosen pairs of particles ${\cal
  N}=\{(i_1,j_1),(i_2,j_2),...(i_M,j_M)\}$. The crosslinks are modelled by
harmonic springs 
\begin{equation}
  \label{Ucross}
  U_{\times} \defeq \frac{1}{a^2} \; \sum_{e=1}^{M} (\vec{R}_{i_e}-
  \vec{R}_{j_e})^2  
\end{equation}
of typical length $a$. (Energies are measured in units of
$k_{\mathrm{B}} T$).  A particular crosslink realisation ${\cal N}$
can be decomposed into $L$ clusters of crosslinked particles:
$\mathcal{N} =\cup_{l=1}^{L} \mathcal{N}_{l}$. The $l$-th cluster,
${\cal N}_l$, contains $N_l$ particles.  All the particles interact
via an isotropic pair potential
\begin{equation}
  \label{pair-pot}
  \UEV \defeq \sum_{1 \le i < j \le N} u(|\vec{R}_{i}- \vec{R}_{j}|)
\end{equation}
which can be taken as a Lennard--Jones potential or just a as hard core
to ensure excluded volume. The details of the potential are irrelevant
for this paper. The total potential energy is the sum of the harmonic
crosslinks and the excluded volume: $ U \defeq U_{\times} + \UEV$ .

\section{Shear viscosity}
We are interested in shear flow and impose an external velocity field
which points in the $x$-direction and increases linearly with $y$:
$\vec{v}(\vec{r}) \defeq\kappa y \vec{e}_x$. Here $\kappa$ denotes the
shear rate which is taken time independent because we aim to compute
the static shear viscosity. In the following we shall need the
symmetrised external flow field $ v_{\mathrm{sym}}^{\nu} \defeq\kappa K^{\nu\mu}
r^{\mu}$ with $K^{\nu\mu} \defeq
(\delta_{\nu,x}\delta_{\mu,y}+\delta_{\nu,y}\delta_{\mu,x})/2$ and
summation over repeated Greek indices implied.  The monomers of the
melt perform a Brownian motion on a semi-microscopic scale. The
monomer configuration at time $t$ is then described by the
$N$-particle probability density
$\psi(\vec{R}_1,\vec{R}_2,...,\vec{R}_N,t)$, which obeys the
time-dependent Fokker--Planck or Smoluchowski equation \cite{DoEd86}
\begin{equation}
  \label{smol-eq}
  \zeta\frac{\partial}{\partial t}\psi
  = \sum_{i=1}^{N}\frac{\partial}{\partial \vec{R}_i}
 \cdot \left(% \frac{1}{\zeta} 
   % \left(
\frac{\partial\psi}{\partial \vec{R}_i}
 + \psi \frac{\partial U}{\partial \vec{R}_i}%\right)
    - \zeta\vec{v}(\vec{R}_i) \psi\right).
\end{equation}
Here $\zeta >0$ denotes the friction constant.

The external velocity field
builds up stresses in the system. Following Kirkwood, they are
described by the stress tensor $\boldsymbol{\sigma}$ with components \cite{DoEd86}
\begin{equation}
  \label{stress}
  \sigma^{\nu\mu}=\rho \;
  \frac{1}{N}\left< \sum_{i=1}^N\frac{\partial 
    U}{\partial R^\nu_i} R_i^\mu\right>_{\!\!\stat}.
\end{equation}
The stress tensor is proportional to the monomer density $\rho$, and
the brackets $\left<\cdot\right>_{\stat}$ indicate the thermal average
over particle positions with respect to the suitably normalised
stationary solution $\psi_{\stat}$ of (\ref{smol-eq}) in the presence
of a shear flow.  Note that the stress tensor as defined in
(\ref{stress}) is {\it symmetric}. The intrinsic linear-response
viscosity is obtained from the stress tensor (\ref{stress}) as $\eta
\defeq \lim_{\kappa\to 0} \sigma^{xy}/(\rho \kappa).$
%\begin{equation}
%  \label{eta0}
%  \eta \defeq \lim_{\kappa\to 0} \frac{\sigma^{xy}}{\rho \kappa}.
%\end{equation}

In order to determine the shear viscosity $\eta$ within linear
response, we follow \cite{Fix83} and expand the stationary probability
density $ \psi_{\stat} = \psi_{\eq}(1+ \kappa \phi) +
\mathcal{O}(\kappa^{2}) $ in powers of the shear rate. The
zeroth-order term is given by the canonical equilibrium state
$\psi_{\eq} \sim \exp\{- U\}$. In order to ensure the correct
normalisation of $\psi_{\stat}$ we require
$\left<\phi\right>_{\eq}=0$, where $\left<\cdot\right>_{\eq}$ denotes
the thermal average over $\{\vec{R}_i\}$ weighted by $\psi_{\eq}$.
The above expansion allows us to express the linear viscosity as an
average over the equilibrium distribution in two alternative forms
\begin{equation}
  \label{eta}
  \eta= \frac{1}{N}\sum_{i=1}^{N} \left< \frac{\partial
      {\phi}}{\partial {R}_i^{\mu}} \, K^{\mu\nu} {R}_i^{\nu}
  \right>_{\!\eq}=\frac{1}{\zeta N}\sum_{i=1}^{N} \left<\frac{\partial
      {\phi}}{\partial \vec{R}_i}\cdot
    \frac{\partial {\phi}}{\partial \vec{R}_i}\right>_{\!\eq}.
\end{equation}
Here only the symmetrised external flow field enters, even if the
externally imposed flow is not symmetric. This is a direct
consequence of the symmetry of the stress tensor as defined in
(\ref{stress}).
  
All the difficulties in the computation of $\eta$ are hidden in the
stationary non-equilibrium distribution $\phi$. In this paper we shall
follow Fixman\cite{Fix83} and derive variational bounds for $\eta$
which circumvent the exact knowledge of $\phi$. Instead a guess for
$\phi$ will serve as a test function in the variational principle with
the quality of the bounds depending strongly on the chosen test
function.  To get a first hint, we consider the simpler case without
excluded-volume interactions (``phantom clusters''), which can be
solved exactly for the viscosity.  In this case the centre of mass
positions of the clusters
\begin{equation} 
  \vec{R}^{(l)}=\frac{1}{N_l}\sum_{i \in \mathcal{N}_{l}} \vec{R}_i
\end{equation}
follow the external flow field, giving rise to a nonzero current in
the stationary state.  The stationary distribution $\psi_{\stat}$
depends only on the particle positions relative to the centre of mass
of the cluster to which they belong, i.e.\ on $\delta \vec{R}_i \defeq
\vec{R}_i-\vec{R}^{(l)}$.  It is explicitly given by $\psi_{\stat}
\sim ( 1+ \kappa \phi^{\times}) \exp\{ - U_{\times}\}$ with
\begin{equation}
  \label{phi-x}
  \phi^{\times} \defeq \frac{\zeta}{2} \; \sum_{l=1}^L \sum_{i \in 
    \mathcal{N}_{l}} \delta {R}_i^{\nu} \, K^{\nu\mu} \, \delta {R}_i^{\mu}. 
\end{equation}
Inserting (\ref{phi-x}) into (\ref{eta}) and using isotropy, we
recover the result
\begin{equation}
  \label{eta-x}
  \eta_{\times}= \frac{\zeta}{2d} \; 
  \sum_{l=1}^{L} \frac{N_{l}}{N} \, \bigl< \Rg^{2}(\mathcal{N}_{l}) \bigr>_{\eq}
\end{equation}
for the viscosity of the Rouse model for gelling polymers
\cite{BrLo99,BrLo01,LoMu05b} in terms of the cluster-weighted sum of
the squared radii of gyration
\begin{equation}
  \Rg^{2}(\mathcal{N}_{l}) \defeq \frac{1}{N_{l}}
  \sum_{i\in\mathcal{N}_{l}}\delta \vec{R}_i ^{2}.
\end{equation}

\section{Bounds for the viscosity}

To derive a {\it lower} bound for the viscosity in the presence of
excluded-volume interactions, we define the functional 
\begin{equation}
  F(\widetilde{\phi}) \defeq 2 \left< \sum_{i=1}^{N}\frac{\partial
      \widetilde{\phi}}{\partial {R}^\nu_i}  K^{\nu \mu}
    {R}_i^{\mu}\right>_{\!\!\eq} 
  -  \frac{1}{\zeta}  \left<  \sum_{i=1}^{N} \frac{\partial
      \widetilde{\phi}}{\partial \vec{R}_i}
    \cdot\frac{\partial \widetilde{\phi}}{\partial \vec{R}_i}\right>_{\!\!\eq}
\end{equation}
for smooth scalar fields $\widetilde{\phi}$. With help of the two
alternative expressions for the viscosity (\ref{eta}), the latter can
be represented as $\eta = F(\phi)/N$.  Functional differentiation of
$F$ with respect to $\widetilde{\phi}$ shows that its global maximum
is attained for $\widetilde{\phi}=\phi$. Hence $ \eta \ge
F(\widetilde{\phi})/N$ for any choice of the test function
$\widetilde{\phi}$.  Choosing the exact solution $\phi^{\times}$ of
the phantom case yields a lower bound for the polymers with excluded
volume
\begin{equation}
  \label{eta-lb-x}
  \eta\ge 
   \frac{\zeta}{2d} \sum_{l=1}^{L} \frac{N_{l}}{N} \,
  \bigl< \Rg^{2}(\mathcal{N}_{l})  \bigr>_{\eq}
\end{equation}
where we used isotropy.

Fixman  \cite{Fix83} has also given an {\it upper} bound for the viscosity
\begin{equation}
  \label{eta-ub}
  \eta \le  \frac{\zeta}{N}\sum_{i=1}^{N}\Bigl< \bigl( K^{\nu\mu}
    {R}_i^{\mu} - {u}_i^{\nu} \bigr) ^{2} \Bigr>_{\eq}
\end{equation}
in terms of $N$ test vector fields $\{\vec{u}_i\}$, $i=1,\ldots, N$,
subject to the divergence condition
\begin{equation}
  \label{u-cond}
  \sum_{i=1}^{N}\frac{\partial}{\partial \vec{R}_i} \cdot
  (\psi_{\eq}\vec{u}_i)=0. 
\end{equation}
The problem with this estimate lies in the difficulty to find test
fields that satisfy condition (\ref{u-cond}). The upper bound
would match the lower bound (\ref{eta-lb-x}) precisely, if we chose
${u}_i^{\nu} = K^{\nu\mu} \delta {R}_i^{\mu}$. However this choice in
general does not fulfil condition (\ref{u-cond}) due to
excluded-volume interactions between monomers in {\it different}
clusters. The above test field is a valid choice for a system which
consists just of a single cluster $\mathcal{N}$. In that case the
upper bound (\ref{eta-ub}) {\it coincides} with the lower bound
(\ref{eta-lb-x}) and the equality
\begin{equation}
  \label{eta-N}
  \eta (\mathcal{N})= 
  \frac{\zeta}{2d} \; \bigl< \Rg^{2}(\mathcal{N}) 
  \bigr>_{\eq,\, \mathrm{s}}
\end{equation}
holds exactly. The additional subscript `s' of the thermal average
indicates that there are no other degrees of freedom besides the
positions of the monomers in the cluster to be averaged---in contrast
to the average in (\ref{eta-lb-x}), which includes all monomers from
all clusters. Eq.\ (\ref{eta-N}) has been suggested on a
phenomenological basis in the context of scaling arguments
\cite{CoGi93,MaAd88,Gen78,Gen79b,Cat85,Mut85,RuCo89,ArSa90,Sa94},
which relate the viscosity to the longest time scale and the latter to
the clusters' linear dimension $\Rg$ and the diffusion constant.

\section{Disorder average over crosslinks}
\label{average}

To obtain the viscosity of the macroscopic polymeric melt, we have to
perform the macroscopic limit $N\to\infty$ and average over different
crosslink realisations. We focus on gelling systems for which the
appropriate crosslink ensemble is given by $d$-dimensional
lattice-bond percolation with bond probability $c/d$, corresponding to
the crosslink concentration $c$.  We denote this combined procedure by
an overbar
$\overline{\phantom{i}\!\boldsymbol\cdot\!\phantom{i}\!}\,$.  For the
viscosity of a polymeric melt, Eq.\ (\ref{eta-lb-x}) yields the lower
bound
\begin{equation}
  \label{eta-result}
  \overline{\eta} \ge  
      %\sum_{n=1}^{\infty} n f(n) \, \eta_{n} \defeq
  \frac{\zeta}{2d} \; \sum_{n=1}^{\infty} n f(n) \, 
   \overline{\langle \Rg^{2}\rangle^{\phantom{i}}_{\eq}}^{\,(n)}
\end{equation}
with $f(n)$ denoting the cluster-size distribution and
$\overline{\phantom{i}\!\boldsymbol\cdot\!\phantom{i}\!}^{\;(n)}\,$
the (normalised) partial average obtained from restricting
$\overline{\phantom{i}\!\boldsymbol\cdot\!\phantom{i}\!}\,$ to
clusters of fixed size $n$. Next we employ the scaling description of
bond percolation for $\varepsilon \defeq |c - \cc| \ll 1$ to compute
critical exponents. In particular, the
percolation average of the radius of gyration of $n$-clusters is known
to obey the scaling relation \cite{StAh94}
\begin{equation}
  \label{r2scal}
   \overline{\langle \Rg^{2}\rangle^{\phantom{i}}_{\eq}}^{\,(n)}
   \Big|_{\varepsilon=0} \sim n^{2/\df}
\end{equation}
with the fractal dimension $\df=(\sigma\nu)^{-1}\approx 2.53$. It is
experimentally well confirmed (e.g.\ in \cite{AdLa91,YiEr98}) and
generally accepted \cite{StCo82,MaAd91} that the scaling
\eqref{r2scal} with the same fractal dimension also applies to
gelation clusters with excluded-volume interactions in a melt.
Moreover, on the theoretical side, thermostatic properties of
vulcanization clusters at criticality were shown to fall into the
universality class of lattice-bond percolation \cite{PeGo00}.  Thus,
when inserting \eqref{r2scal} into (\ref{eta-result}), we obtain that
the averaged viscosity of the polydisperse melt diverges with a
critical exponent
\begin{equation}
  \label{k-result}
  k \ge  2\nu -\beta
\end{equation}
at the sol-gel transition. We recall that throughout our calculations
we have not made any assumptions on the detailed nature of the
excluded-volume potential. 

On the other hand, if excluded-volume interactions were absent, the
geometric fractal dimension $\df$ would take on the Gaussian or ideal
Rouse value $\df^{(\mathrm{G})} \approx 4$. The resulting lower bound
for $k$ would then coincide with the exact result in the phantom case that was
computed previously \cite{BrLo99,BrLo01,LoMu05b}.

Finally, we turn to the viscosity (\ref{eta-N}) of a single cluster
with excluded-volume interactions that has been isolated from a
gelling melt. On average, we get
\begin{equation}
  \label{eta-Nav}
  \eta_{n,\,\mathrm{s}} \defeq \overline{\eta(\mathcal{N})}^{\,(n)} 
   =  \frac{\zeta}{2d} \;
  \overline{ \bigl< \Rg^{2}(\mathcal{N})  \bigr>_{\eq,\, \mathrm{s}}}^{\,(n)},
\end{equation}
which is an exact result. For single clusters isolated from \emph{critical}
melts, (\ref{eta-Nav}) implies the Mark--Houwink relation
$\eta_{n,\,\mathrm{s}}|_{\varepsilon=0} \sim n^{2/\dfs}$ with the
fractal dimension $\dfs$ for a single cluster.  Whereas for clusters
in the melt the excluded volume interaction is (partially) screened,
this is not the case for an isolated cluster, resulting in a lower value
of $\dfs$ as compared to $\df$.  A generalisation of Flory's argument
to clusters \cite{Lubensky} suggests $\dfs =2$.  This exponent value
is observed in light-scattering experiments for solutions obtained
from diluting a polydisperse melt \cite{MaKe86,AdLa97,LeNi99}.

%%%%%%%%%%%%%%%%%%%%%%%%%%%%%%%%%%%%%%%%%%%%%%%%%%%%%%%%%%%%%%%%%%%%%%%%%%
%
\section{Discussion and outlook}
%
%%%%%%%%%%%%%%%%%%%%%%%%%%%%%%%%%%%%%%%%%%%%%%%%%%%%%%%%%%%%%%%%%%%%%%%%%%

The exact lower bound for the critical exponent is definitely in
disagreement with several entries in Table~\ref{data}. For simulations
or experiments referring to a melt, the only way out are effects due
to entanglements (besides crossover effects due to a tiny critical
region). Entanglement effects cannot be ruled out, but are generally not
expected to play a vital role on the sol side of the transition --
quite in contrast to the well crosslinked gel phase, where
entanglement effects due to interlocking loops are clearly
important\cite{EvKr95, EvKr96}.

As far as dense solutions are concerned, one needs to consider
hydrodynamic interactions due to the presence of a solvent.  It seems
feasible to include pre-averaged hydrodynamic interactions in the
variational approach pursued here. This may provide a semi-microscopic
starting point for the computation of upper and lower bounds for
viscosities for solutions of gelling polymers. Such bounds are
all the more interesting, because a dense solution of gelling
\emph{phantom} polymers is known to behave completely unphysically,
showing a finite viscosity at the sol-gel transition \cite{LoMu05a}.

%%%%%%%%%%%%%%%%%%%%%%%%%%%%%%%%%%%%%%%%%%%%%%%%%%%%%%%%%%%%%%%%%%%%%%%%

\acknowledgments
We acknowledge financial support by the DFG under grant Zi209/7-2 and
through SFB~602. 

%%%%%%%%%%%%%%%%%%%%%%%%%%%%%%%%%%%%%%%%%%%%%%%%%%%%%%%%%%%%%%%%%%%%%%%%%

\end{document}